# A Modified Mesh with Individually Monitored Interferometers for Fast Programmable Optical Processors

**Kaveh (Hassan) Rahbardar Mojaver\*, Bokun Zhao, and Odile Liboiron-Ladouceur**
*Department of Electrical and Computer Engineering, McGill University, Montréal, QC, H3E 0E9, Canada*
*\*hassan.rahbardarmojaver@mcgill.ca*

**Abstract:** We demonstrate a novel mesh of interferometers for programmable optical processors. Employing an efficient programming scheme, the proposed architecture improves energy efficiency by 83% maintaining the same computation accuracy for weight matrix changes at 2 kHz. © 2022 The Author(s).

1. **Introduction**

Programmable optical processors are promising structures for ultrafast classic and quantum computations. These processors can be based on a mesh of Mach-Zehnder interferometers (MZIs) and can perform vector matrix multiplication widely required in machine learning applications [1]. To perform a specific vector matrix multiplication, we must first program the optical processor. The process of programming sets all the MZIs' bias towards achieving a defined weight matrix. In *in-situ* programming, the MZIs biases necessitate time/energy consuming optimization techniques such as back propagation or gradient descent [2]. In *ex-situ* programming, one calculates the bias points required for a specific weight matrix using an external electronic processor and implements them on the optical processor. In both techniques, every time the weight matrix changes, the processor must be reprogrammed. The time-consuming in-situ programming limits the application of optical processors to stationary weight matrix tasks or tasks with low frequency variation in the weight matrix. The downside of ex-situ programming is that once the calculated weight matrix is applied to the phase shifters, dynamic errors such as thermal crosstalk between phase shifters or small drift in the bias points set by the electronic drivers degrade the accuracy of the processor. In such situation, the phase shifters' bias must be readjusted to compensate for those changes.

The processor programming and its robustness is faster through a closed loop control system for monitoring the phase setting of each phase shifter. However, solutions such as waveguide taps and in-line photodetectors often increase the insertion loss of the structure limiting the scalability [3]. Therefore, architectures ideally require monitoring the phase shift applied by a specific phase shifter within the optical paths without modifying the bias of the other phase shifters to mitigate thermal crosstalk. The architecture we present in this work, referred as Bokun mesh, provides phase monitoring while maintaining the minimum optical path depth. Bokun mesh is a topology arrangement that merges the attributes of two mesh topologies. With in-situ monitoring, the total energy efficiency improves by 83% compared to the rectangular mesh for a 10 × 10 mesh with weight matrix changing at 2 kHz. The performance of Bokun mesh enabled by an optimal optical depth is three times more resilient to the loss and fabrication imperfections compared to the Reck and Diamond topologies [4, 5] for a 10 × 10 mesh used in a two-layered optical neural network for MNIST classification task.

2. **Phase Monitoring in Optical Processor**

An N × N multiport reconfigurable MZI-based optical processor is a unitary optical component which consists of $n$ MZIs connected to each other based on a given mesh topology. The Reck mesh topology theorized by Reck et al. consists of a triangular mesh of MZIs [4]. Reck mesh can be employed to implement an arbitrary N × N unitary matrix with $n = N(N-1)/2$ MZIs connected to each other. The triangular shape of the Reck mesh enables sequential calibration starting from the MZI in the vertex and then through the rest of the structure. The Diamond mesh proposed by our group (Shokraneh et al.) employs a total MZI of $(N-1)^2$ which includes additional MZIs coping with loss imbalance due to an increased mesh symmetry compared to the Reck topology [5]. It also provides diagonal optical path useful for monitoring and programming. The rectangular mesh proposed by Clements et al. uses the same number of MZIs as the Reck mesh [6]. As input signals mix with adjacent signals earlier in the Clements topology, the shorter optical depth leads to lower loss compared to the Reck topology in which the bottom input signals propagate over a relative longer distance before interacting with other signals. However, the Clements structure is not triangular as Reck and Diamond, which makes its calibration more complicated because in calibrating each MZI of the structure, the light passes through non-calibrated MZIs affecting the overall process. In this work, we propose the Bokun mesh topology offering minimum optical depth while maintaining the Diamond shape structure essential to the fast and energy efficient programming. Indeed, the Bokun mesh is a truncated Diamond mesh with the middle optical inputs/outputs (I/Os) used as the main optical path and the peripheral I/Os for calibration purpose only. It can also be considered as an extended version of Clements mesh with those additional MZIs at the top and bottom of the structure providing diagonal I/O paths for each individual MZI.

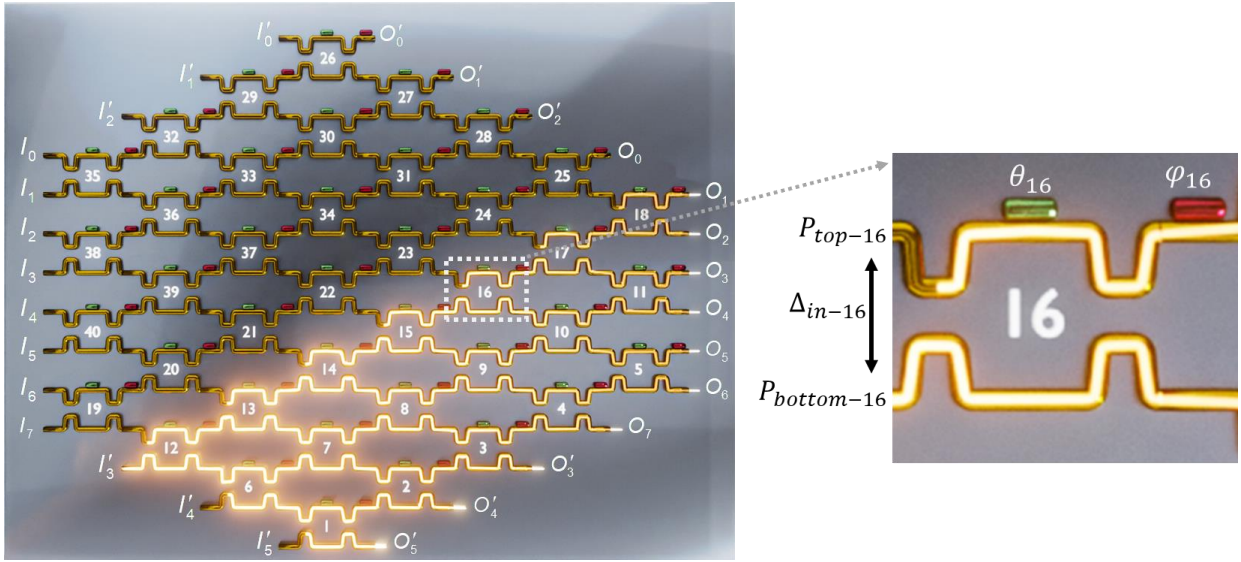

Fig. 1. 8 × 8 Bokun mesh topology with 40 MZIs ($n = N(N + n/2 - 2)/2$). $I_0$ to $I_7$ and $O_0$ to $O_7$ are the optical I/Os of the signal paths. $I'_0$ to $I'_5$ and $O'_0$ to $O'_5$ provide the diagonal paths for monitoring and calibration. The illuminated section of the mesh highlights the monitoring of phase setting of MZI-16 by applying light only into $I'_3$ and detecting the light at $O_1$.

The linear transformation matrix of the 2 × 2 MZI building block is shown below, for a fixed state of polarization, 50:50 splitting ratio of the two couplers, and assuming lossless optical propagation.

$$\begin{bmatrix} O_{top} \\ O_{bottom} \end{bmatrix} = e^{j(\theta/2)} \begin{bmatrix} e^{j\varphi} \sin\left(\frac{\theta}{2}\right) & e^{j\varphi} \cos\left(\frac{\theta}{2}\right) \\ \cos\left(\frac{\theta}{2}\right) & -\sin\left(\frac{\theta}{2}\right) \end{bmatrix} \begin{bmatrix} I_{top} \\ I_{bottom} \end{bmatrix}, \qquad \text{eq. (1)}$$

where $\theta$ is the internal phase shift changing the output optical intensity, and $\varphi$ is the external phase shift defining the output optical phase. The optical electric fields of a plane wave at the input and output ports of a given MZI are labelled as $I_{top}$ and $I_{bottom}$, $O_{top}$, and $O_{bottom}$, respectively. Now, the question is in which condition we can monitor the state of an MZI phase shifters (for example MZI-16) when multiple MZIs are cascaded. To find the answer, we define the cross state transmission of MZI-16, i.e., the transmission from its bottom input to its top output.

$$\left(\cos\left(\frac{\theta_{16}}{2}\right) + \sin\left(\frac{\theta_{16}}{2}\right) \sqrt{\frac{P_{top-16}}{P_{bottom-16}}} e^{-j\Delta_{in-16}}\right)^2, \qquad \text{eq. (2)}$$

in which $P_{top-16}$ and $P_{bottom-16}$ are the optical signal powers at the top and bottom input ports of MZI-16 with a relative phase shift difference of $\Delta_{in-16}$. As the top port of MZI-16 has no optical light, the transmission is proportional to $\cos^2(\theta_{16}/2)$ such that we can monitor the state of $\theta_{16}$ from the transmission. We then analyze the effect of the subsequent MZIs on MZI-16 in the optical path towards the output detector. In this case, the transmission of MZI-17 is added to obtain the transmission from the bottom input of MZI-16 to the top output of MZI-17.

$$\cos^2\left(\frac{\theta_{16}}{2}\right) \cdot \left(\cos\left(\frac{\theta_{17}}{2}\right) + \sin\left(\frac{\theta_{17}}{2}\right) \sqrt{\frac{P_{top-17}}{P_{bottom-17}}} e^{-j\Delta_{in-17}}\right)^2, \qquad \text{eq. (3)}$$

in which $P_{top-17}$ and $P_{bottom-17}$ are the optical signal powers at the top and bottom input ports of MZI-17 with a relative phase shift difference of $\Delta_{in-17}$. For a given $\theta_{17}$ when $P_{top-17} = 0$, the impact of MZI-17 does not affect MZI-16. Similarly, the transmission through MZI-18 can be added in a way to enable MZI-16 only by nulling the following MZIs in the calibration optical path.

In the Bokun mesh, the phase monitoring is possible for all MZIs. Note that the Diamond mesh also provides this feature since there is a diagonal path from an input to an output passing through each MZI. The first main advantage of Bokun mesh compared to Diamond mesh is employing the center I/Os which allows each input signal to mix with other input signals earlier in the structure leading to shorter optical depth. The second advantage is mitigating the additional MZIs at the input and output of the structure minimizing the optical depth while maintaining the peripheral MZIs for calibration and monitoring.

3. Performance Evaluation

Figure 2a presents the energy consumption of four 10 × 10 different optical processor structures: the Reck, Diamond, Clements, and Bokun meshes. We considered a power dissipation of 20 mW/π for the thermo-optic phase shifters (TOPS) on silicon on insulator (SOI) platform [7, 8]. When the weight matrix is static, the power consumption is essentially for computation. In such scenario, Reck and Clements structures with a smaller number of MZIs show better efficiency. For dynamic changes in the weight matrix, the energy efficiency deteriorates due to the programming procedure which is often neglected in reported studies. For Reck and Clements, we assumed a back propagation programming method with 200 iterations [2]. The ex-situ programming (i.e., pre-defined weight matrix programming) is not viable on these meshes due to the lack of monitoring options. For Diamond and Bokun, we considered ex-situ programming with 10 iterations/MZI monitoring the MZI state and readjusting their bias. The power consumption required by the electronics is not accounted in both the back propagation and ex-situ programming. The programming time is estimated based on a 2.2 µs transit time in TOPS [7].

As shown in Fig. 2b, the back propagation considerably increases the energy consumption of the processor, however, the two architectures with monitoring options, can provide relatively fast programming, keeping the energy efficiency relatively unchanged. In the case of Clements, the energy consumption increases from 450 fJ/Op (Joules per operation) for stationary weight matrix to 3750 fJ/Op for weight matrix changing at 2 kHz. While for Bokun, it increases only from 610 fJ/Op to 638 fJ/Op. To evaluate the meshes performance in presence of fabrication imperfections, we compared the classification accuracy of a 2-layered MNIST classifier based on the four presented architectures (Fig. 2d-g) [9]. A figure of merit (FOM) in the unit of dB.rad defines the region where the classification accuracy is higher than 75%. The Clements and Bokun with minimum optical depth provide more robustness against optical loss, i.e., 16.9 dB.Rad and 15.8 dB.Rad, respectively (Fig. 2c). However, the classification accuracy in Reck and Diamond decreases with insertion loss leading to a reduction in the FOM, i.e., 5.0 dB.rad and 5.1 dB.rad, respectively.

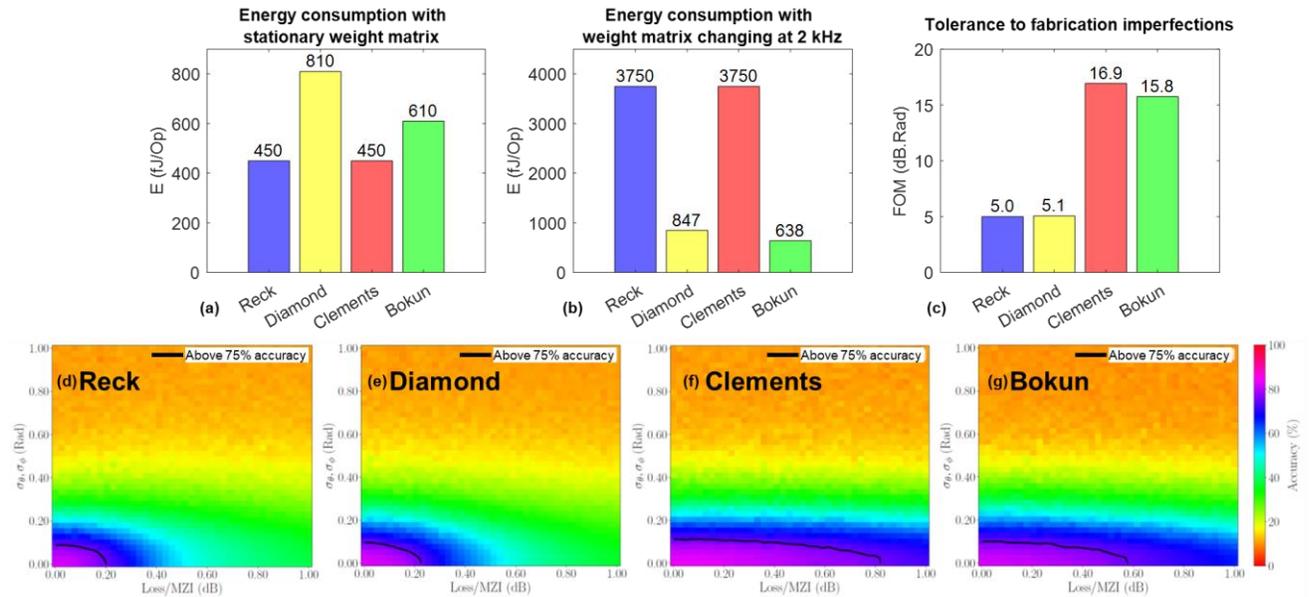

Fig. 2. Energy consumption in units of energy per operation (a) with and (b) without programming; (c) FOM in units of dB.Rad for fabrication imperfection tolerance while running MNIST task; (d)-(g) contours of classification accuracy in the presence of loss and phase error outlining a 75% accuracy.